\documentclass[twocolumn,showpacs,preprintnumbers,amsmath,amssymb,prb,floatfix]{revtex4}

\usepackage{graphicx}
\usepackage{dcolumn}
\usepackage{bm}
\usepackage{epstopdf}
\begin{document}


\title{Strain and Electric Field Modulation of the Electronic Structure of Bilayer Graphene }

\author{B. R. K. Nanda}
\author{S. Satpathy}%
\affiliation{%
Department of Physics $\&$ Astronomy, University of Missouri,
Columbia, MO 65211}%

\date{\today}

\begin{abstract}
We study how  the electronic structure of the bilayer graphene (BLG) is changed by electric field and strain from {\it ab initio}
density-functional calculations using the LMTO and the LAPW methods. 
Both hexagonal and Bernal stacked structures are considered.
The BLG is a zero-gap semiconductor like the isolated layer of
graphene. We find that while strain alone does not produce a gap in
the BLG, an electric field does so in the Bernal structure but not in
the hexagonal structure. The topology of the bands leads to Dirac
circles with linear dispersion in the case of the hexagonally stacked
BLG due to the interpenetration of the Dirac cones, while for the
Bernal stacking, the dispersion is quadratic. The size of the Dirac
circle increases with the applied
electric field, leading to an interesting way of controlling the
Fermi surface. 
The external electric field is screened due
to polarization charges between the layers, leading to a reduced size of the band gap and the Dirac circle. 
The screening is substantial in
both cases and diverges for the Bernal structure for small fields as
has been noted by earlier authors. As a biproduct of this work, we present the tight-binding parameters for the free-standing single layer graphene as obtained by fitting to the density-functional bands, both with and without the slope constraint for the Dirac cone.
\end{abstract}
\maketitle
\section{Introduction}

\begin{figure}
\includegraphics[width=9cm]{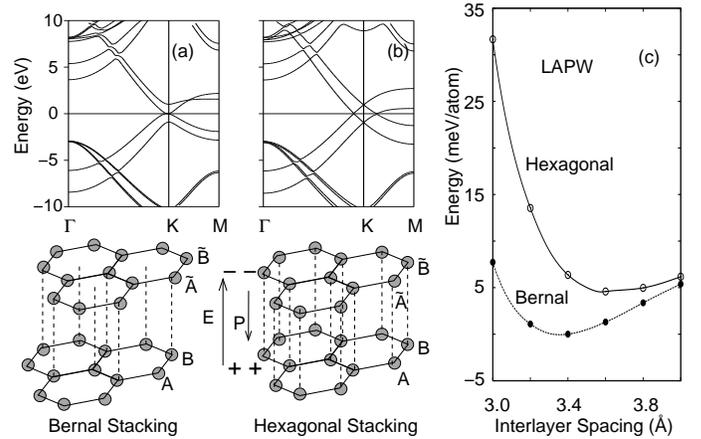}
\caption{\label{introfig} Band structure of Bernal stacked (a) and
hexagonal stacked (b) BLG
without any strain or electric field as
obtained from the LMTO calculations. Fig. (c) shows the computed
total energy per carbon atom as a function of the interlayer spacing. }
\end{figure}

Recently bilayer graphene (BLG) has been shown to possess a band gap in
the presence of an external electric field,\cite{ohta,oostinga} an
effect that has important implications
for transport across the graphene layers and for possible device applications.
While the freestanding BLG
is a zero band gap semiconductor\cite{latil,mccann1} like the single
layer graphene (SLG),
an applied electric field through an external gate induces an
asymmetric potential between the graphene layers, which in turn
opens a gap between the valence and the conduction bands in the
Bernal structure.\cite{mccann1,mccann2,castro,min,nilsson}
It has been recently noted\cite{liu} that graphene exhibits the hexagonal stacking surprisingly often, surprising because of the higher energy of the hexagonal structure.  In the hexagonal
structure, the two graphene layers are stacked vertically on top of
one another, while in the Bernal strcuture, one layer is rotated with
respect to the other as indicated in  Fig. \ref{introfig}, so that
half of the atoms are directly above the carbon atoms in the first
layer, while the remaining half lie on top of the hexagon centers.
The difference in symmetry between the Bernal and the hexagonal
structures leads to substantial differences in the band structure,
especially, in the formation of a band gap, an effect we study in this paper.
%

To tune the field-induced band gap of the BLGs for practical
applications in nano-electronic devices, it is important to examine
the
factors having strong influence on its electronic structure.
Electronic structure of BLG was shown to be sensitive to the
interlayer spacing due to coupling between the graphene
layers\cite{ohta1, varchon}. Hence, a uniaxial strain, which changes
the interlayer spacing, could modulate the field induced band
gap\cite{guo,raza}.
The other factor that controls the gap is the screening of the
electric field. An applied electric field
causes unequal charge distribution among the graphene layers which in
turn creates a polarized electric field in the
opposite direction,\cite{min} thereby reducing the effective electric field. A detail investigation through
density-functional calculations is required to understand the effect
of the electric field and strain on the electronic structure. 

While the electronic structure of the Bernal BLG has been theoretically examined, 
such studies have not been performed for the hexagonal structure to our knowledge. 
In this paper, we report results from density-functional calculations
and tight-binding models to examine the modification of electronic
structure of both Bernal and hexagonal stacked BLG by electric field
and strain. In the Bernal structure, the screening of the external
electric field substantially reduces the band gap as shown earlier,
while in the hexagonal structure the screening is not as strong, as
explained from a simple tight-binding model. In the Bernal structure,
where a gap can be opened up by the electric field, strain can be
used in addition to modify its magnitude. We find that the gap can be
increased considerably by reducing the interlayer spacing from its
equilibrium value by a small amount. In the hexagonal structure, the
gap cannot be opened up; however, an interesting point for this
structure is that the topology of the bands leads to Dirac circles
centered about the K point in the Brillouin zone with linear
dispersion due to interpentration of Dirac cones. Though the electric
field does not produce a gap, it increases the size of the Dirac
circles leading to a novel way of controlling the Fermi
surface.

Density functional calculations presented in this
paper
were performed using the Linear
Muffin-Tin Orbital (LMTO) Method \cite{lmto} and linear augmented
plane wave (LAPW) method \cite{wien} using the local density
approximation (LDA). While the LAPW method was used to obtained the total energies and the optimized
structures, the electronic structure
was investigated using the LMTO method, which did not produce any substantial difference as far as the band structure is concerned. 
To study the field
modulated band structure of the BLG, an extra potential $\Delta$ was
added in the LMTO method to the on-site energies of the carbon s and p orbitals of one
individual layer.  We used the periodic
boundary condition normal to the BLG planes, keeping about 10 \AA \
of vacuum between the different BLG layers  and, furthermore, in the
LMTO method, we included layers of empty spheres at positions
compatible with the symmetry of the particular BLG structure. The
symmetry compatibility was especially important for studying the
screening effects at low electric fields.

\section{Tight-Binding parameters for a single graphene monolayer}

Before we present the results for the BLG, in this section, we
present the tight-binding fitting parameters for the monolayer
graphene obtained by fitting with the LAPW bands. These parameters
will be used in a tight banding description of the BLG.

\begin{figure}
\includegraphics[width=8cm]{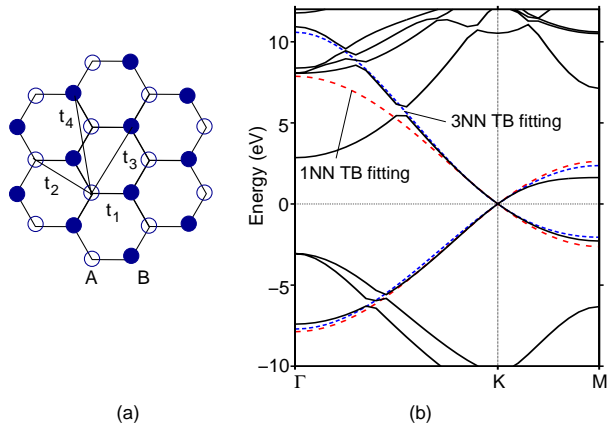}
\caption{(color online) Tight-binding fitting (red and blue lines) of
the LAPW bands (black lines) for a monolayer graphene and without the
slope constraint at the Dirac point (see text). Tight-binding fitting
was made for the p$_z$ bands by retaining up to four nearest
neighbors and the fitting parameters are presented in Table 1.  While
just the first NN hopping is enough to fit the slope of the Dirac
cone at the K point, at least three NNs must be retained to describe
the band structure well in the entire Brillouin zone. }
\label{tbfit}
\end{figure}

For the freestanding monolayer graphene, it is known
that the formation of the Dirac cones at the Fermi
surface is the outcome of the p$_z$-p$_z$ $\pi$ hopping between the two inequivalent carbon atoms A and B in the unit cell.
Though a simple tight-binding model involving the p$_z$-p$_z$
interaction within the nearest neighbor (NN) coordination explains
the linear band dispersion at the Dirac points K and K$^\prime$, it
is essential to include further neighbor hoppings if an accurate
description of the band structure in the entire Brillouin zone is
desired. We find that for this, al least three NN hopping integrals
must be retained as indicated from the root-mean-square deviation given in Table 1 and the band structure of Fig. \ref{tbfit}.

The TB Hamiltonian is a 2 $\times$ 2 matrix
\begin{eqnarray}
{\cal H}  =\left( \begin{array}{cc}
h_{AA}&h_{AB}\\
h_{AB}^*&h_{BB}\\
\end{array}\right) 
\end{eqnarray}
where A and B denote the two carbon atoms in the unit cell. Retaining
only the first two NN hoppings, the form of the matrix elements are
\begin{eqnarray}
h_{AA}  &=& t_2 [2 \cos (\sqrt3k_xa)+4 \cos(\sqrt3k_xa/2)
\cos(3k_ya/2)],\nonumber\\
h_{AB} &=& t_1[2 \exp (-ik_ya/2) \cos(\sqrt 3 k_xa/2)+ \exp(ik_ya)],
\nonumber\\
\end {eqnarray}
and $h_{AA} = h_{BB}$. This can be easily generalized to further near neighbors. The symbols
$t_1$, $t_2$, $t_3$, and $t_4$ are the NN hopping parameters as shown
in Fig. \ref{tbfit}, `a' is the C-C bond length, and the on-site energy of the carbon orbital is taken to be zero. The slope of the Dirac cone centered at the K or the K$^\prime$ point, easily obtained by diagonalizing the Hamiltonian and taking the limits, is
given by the expression
\begin{equation}
v \equiv [ d\epsilon_k / dk ]_K = a (3t_1/2 - 3 t_3 + 9t_4/4). 
\label{slope}
\end{equation}

The slope at the Dirac point is about $5.4 eV \cdot$ \AA, as calculated from LAPW, which corresponds to the velocity of $8.2 \times 10^5$ m/sec. 
The TB bands
were fitted to the LAPW bands by keeping hopping integrals for a
certain number of NNs and neglecting the hopping for further
neighbors. We obtained two sets of optimized  parameters. The first
set of TB parameters was obatined by constraining the slope of the
linear bands at the Dirac point to the LAPW value,  while the second set was obtained
without this constraint. For low energy properties, where the magnitude of the slope may be important, the second set of the parameters should be used.
The results are shown in Fig. \ref{tbfit} and Table I. In our TB work
of the BLG below, we have retained only the first NN interaction with
the slope constraint, since this is sufficient for our purpose as we are mainly interested in states close to the
band gap region.

\begin{table}
\caption{Tight-binding NN hopping integrals obtained from the least
square fitting of the LAPW bands.  Fitting was done both with and without constraining the
slope of the Dirac cone to its LAPW value. Quality of the fit is indicated by the
root-mean-square deviation over the entire Brillouin zone.}
\begin{center}
\begin{tabular}{c|ccccc}
\hline
Range of &t$_1$&t$_2$&t$_3$&t$_4$&RMS\\
hopping&&(eV)&&&deviation\\
\hline
\multicolumn{5}{c}{Without Slope Constraint}\\
\hline
1NN&-2.625&&&&0.42\\
2NN&-2.910&0.160&&&0.22\\
3NN&-2.840&0.170&-0.210&&0.11\\
4NN&-2.855&0.170&-0.210&0.105&0.10\\
\hline
\multicolumn{5}{c}{With Slope Constraint}\\
\hline
1NN&-2.56&&&&0.46\\
2NN&-2.56&0.160&&&0.32\\
3NN&-2.90&0.175&-0.155&&0.12\\
4NN&-2.91&0.170&-0.155&0.02&0.115\\
\hline
\end{tabular}
\end{center}
\end{table}

\section{Bilayer Graphene in the Hexagonal structure}
\begin{figure}
\includegraphics[width=6cm]{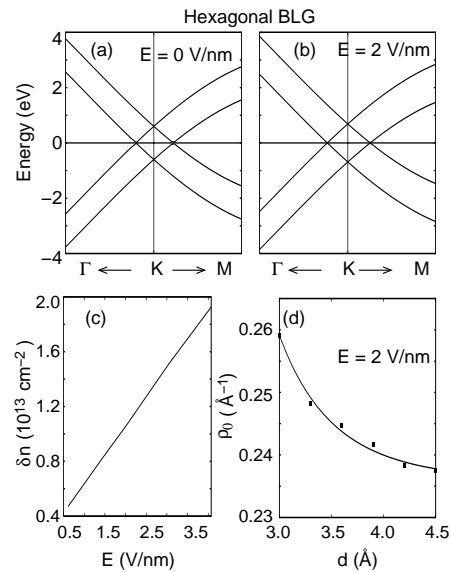}
\caption{\label{aaband} Band structure of the hexagonal
stacked BLG without (a) and with (b) an electric field and the variation of the radius of the Dirac circle (c).
The charge density difference $\delta n$ between the layers induced 
by the  electric field is shown in (c). All results were obtained from the density-functional LMTO calculations.}
\end{figure}

\begin{figure}
\includegraphics[width=7.5cm]{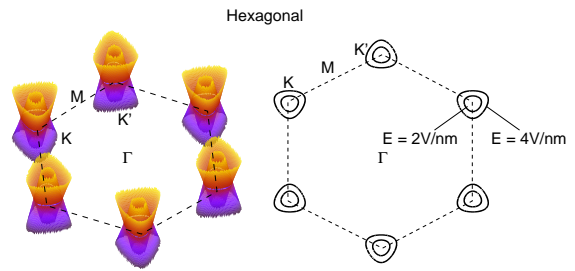}
\caption{\label{pretty}(color online) Topology of the band structures of
hexagonal   BLG with equilibrium interlayer separation and
in the presence of an electric field. The Fermi
surface is formed by two interpenetrating cones, forming a
``Dirac" circle centered around the K and the K$^\prime$ points in the Brillouin zone. The size of the circle can be
controlled by the electric field as the right hand figure indicates for two different electric fields.
 }
\end{figure}

As mentioned already, recent experiments suggest that the hexagonal BLG
is surprisingly common\cite{horiuchi} in spite of its higher energy as obtained from the total energy calculation 
(Fig. \ref{introfig} and Ref. 17) that indicate the Bernal BLG to be higher than the hexagonal BLG by $\sim$ 5 meV/C atom. 
 For the
equilibrium interlayer spacing (d = 3.51 \AA ), the band structure for the hexaoganal stacked BLG is shown in Fig.
\ref{aaband}(a) and the topology of the bands near the Fermi surface
is shown in Fig. \ref{bandsketchaa}. Unlike the case of the Bernal stacked BLG, here we
see two interpenetrating Dirac cones, which intersect to form a circular Fermi surface, the Dirac circle. 

      An electric field does not change the overall feature
of the band structure (see
Fig. \ref{aaband}(b)) and in contrast to the Bernal BLG, it does not produce a band gap because of the
higher symmetry situation in the hexagonal stacking. 
The main features of the band structure can be understood from a tight-binding
(TB) model involving the nearest-neighbor $\pi$- interaction  of the
p$_z$ orbitals  in the graphene plane and the $\sigma$- interaction
between the planes. With the four inequivalent carbon sites,
one can form the Hamiltonian as in the monolayer case and in addition
add the interlayer coupling term. One can then expand the Hamiltonian
matrix elements for a
small momentum around the Dirac point, which
then becomes an excellent approximation for the band gap region. The
result is
\begin{eqnarray}
{\cal H}_{hex} =\left( \begin{array}{cccc}
\frac{\Delta}{2}&t&0&\nu\pi^{\dagger}\\
t&-\frac{\Delta}{2}&\nu\pi^{\dagger}&0\\
0&\nu\pi&-\frac{\Delta}{2}&t\\
\nu\pi&0&t&\frac{\Delta}{2}\\
\end{array}\right),
\label{HAA}
\end{eqnarray}
where $\nu = \frac{3}{2}t_{1}a$ is the Fermi velocity
for the monolayer,
$\pi=k_x + ik_y$ is the complex momentum with
respect to the Dirac point K,
$\Delta$ is the potential difference
between the two layers, $t$ is the interlayer hopping integral ($  
\sim 0.6 eV$ as extracted from the LAPW results), and the
basis set of the Hamiltonian
consists of the Bloch functions made out of the 
carbon orbitals
located at A, $\tilde{A}$, $\tilde{B}$, and B,
 respectively and  in that order. The atoms are indicated in Fig. \ref{introfig}. The electric field potential $\Delta$ appearing in the Hamiltonian Eq. \ref{HAA} corresponds to the final screened field experienced by the electrons and not to the bare electric field which is reduced by the polarization field due to screening, so that
$\Delta = \Delta_{ext}/\epsilon$, where $\Delta_{ext}$ is the external electric field and $\epsilon$ is the static dielectric constant.

The Hamiltonian is easily diagonalized to yield the eigenvalues
\begin{equation}
\varepsilon(k) =  \pm\xi \pm\nu k,
\label{bandaa}
\end{equation}
where 
\begin{equation}
  \xi = \sqrt{\Delta^2/4 + t^2} 
\end{equation}
  specifies the energy locations of the vertices of the Dirac cones. The energy dispersion is sketched in Fig. \ref{bandsketchaa}, which shows the two Dirac cones with linear dispersion with the vertices
of the cones separated by the energy $2 \xi$. 

The radius of the
circular Fermi surface, the ``Dirac" circle, where the cones intersect the zero of energy, is simply
$
\rho_0 = \xi /v.
$
Since for small fields the screened potential $\Delta$ is proportional to the 
applied field, the equation indicates that $\rho_0$ increases both with the applied electric field
and the interlayer hopping parameter ($t$ increases as the interlayer separation decreases). 
This is a notable feature of the hexagonal BLG, viz., that we have a circular Fermi surface
with zero band gap and that the radius of the circle can be modified by electric field and strain. Any doped holes or electrons will form a thin shell in the momentum space 
around the Dirac circle. This may lead to interesting electronic and
magnetic properties in the presence of impurities.

\section{Bilayer Graphene in the Bernal Structure} 

\begin{figure}
\includegraphics[width=8cm]{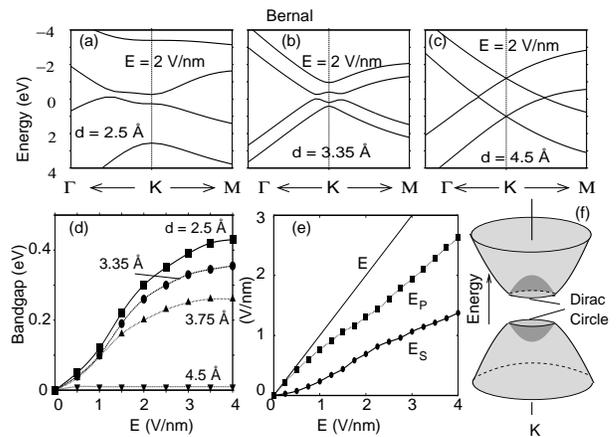}
\caption{\label{bernalefield} Electric field modulated band structure
of the Bernal stacked BLG   for different interlayer spacings (a - c) and the
variation of the band gap as a function of strain and electric field
(d).  Fig. (e) shows the magnitude of the
polarization field $E_p$ and the net screened electric field $E_s$ as a function of the applied electric field $E$. All results were obtained from the density-functional LMTO calculations. Topology of the bands in the gap region is sketched in Fig. (f). 
}
\end{figure}

The effect of strain and electric field on the band structure of the Bernal BLG
is summarized in Fig. \ref{bernalefield}. 
As seen from  Fig. \ref{introfig}(a), for the Bernal BLG, there are four nearly
parabolic bands near
the Fermi energy and two of these touch at the Dirac point making it
a zero band-gap semiconductor. However, in the presence of an electric
     field, these two bands split opening up a small gap
(Fig. \ref{bernalefield}(b)) consistent with earlier density functional
results\cite{guo,raza}. The magnitude of the band gap
increases linearly for small electric fields and saturates
for higher electric fields as seen from Fig. \ref{bernalefield}(d).
 
For
large interlayer spacing, there is very little coupling between the
two individual sheets and the bands for the single graphene sheet are
reproduced (Fig. \ref{bernalefield}(d)). If the interlayer separation $d$ is large enough ($>$ 4.5 \AA ), the gap
vanishes irrespective of the electric field and  
decreasing the magnitude of $d$ increases the gap substantially. Quite interestingly, if $d$ is too small ($ d <$ 2.7 \AA ), the band gap
becomes indirect due to interactions with other orbitals in addition
to p$_z$.
However, the indirect band gap may not be realized
in practice because of the large strain needed, which may also lead
 lead to a possible deformation of the
graphene
sheets. For reasonable layer separations, the electric field produces a small direct gap at k points,
the locus of which is nearly a circle (the ``Dirac" circle) centered
around   the K or the K$^\prime$ points in the Brillouin zone as indicated from the band structure of
Fig.  \ref{bernalefield}(b) and Fig. \ref{pretty}.

The main features of the band structure can again be explained by a nearest-neighbor TB model, analogous to the case for the hexagonal structure. The resulting Hamiltonian is
\begin{eqnarray}
{\cal H}_{Bernal} =\left( \begin{array}{cccc}
\frac{\Delta}{2}&0&0&\nu\pi^{\dagger}\\
0&-\frac{\Delta}{2}&\nu\pi&0\\
0&\nu\pi^{\dagger}&-\frac{\Delta}{2}&t\\
\nu\pi&0&t&\frac{\Delta}{2}\\
\end{array}\right),
\label{HAB}
\end{eqnarray}
where the
basis set of the Hamiltonian
consists of the Bloch functions made out of the
carbon orbitals
located at A, $\tilde{B}$, $\tilde{A}$, and B, in that order.
Diagonalization of the Hamiltonian yields the four eigenvalues
\begin{eqnarray}
\epsilon(k)&=&  \pm\frac{1}{2} [  \Delta^2 +
4\nu^{2}\pi\pi^{\dagger}+ 2t^{2} \nonumber\\
&& \pm
2 (  4\nu^{2}\pi\pi^{\dagger}(\Delta^2+t^2) + t^4)^{1/2} ]^{1/2}.
\label{bandab}
\end{eqnarray}
At the K or the K$^\prime$ points in the Brillouin zone, the energies of the bands are $\epsilon = 
\pm\frac{\Delta}{2}, \pm 
\sqrt{\frac{\Delta^2}{4} + t^2}$, and they disperse quadratically
as seen from
Fig. \ref{bernalefield}(b) such that a gap of magnitude 
$E_g = 
\Delta t /(\Delta^2+t^2)^{1/2}$ is produced at k points in the 
Brillouin zone that make the ``Dirac circle" of radius $\rho_0 = 
\Delta / 2\nu$.
The above expression clearly shows that E$_g$ 
increases linearly
for low fields $(\Delta << t)$ and saturates at
high fields ($\Delta >> t$). This is consistent with the density-functional results
for the dependence of the band gap on the external field $E$
and bilayer spacing shown in 
Fig. \ref{bernalefield}(d).

\section{\label{screening}Screening in the BLG}
It has already been pointed out that the screening is divergent in the Bernal structure for small applied fields between the layers,\cite{mccann1,min} which is a consequence of the band topology at low energies. In this Section, we examine the screening in the hexagonal structure and compare with the results for the Bernal structure.  An applied electric field $E$ transfers electrons from one graphene layer to the other, which produces a polarization field $E_p$ resulting in the net screened field $E_s$ that the electrons see, so that
 we have
$E_s = E - E_p$.

We compute the screening effects in three different ways  using (a) density-functional LMTO method, (b) the tight-binding model, and finally (c) from an analytic expression obtained by using the linear band dispersion. Unlike the Bernal structure, we find that the screening does not diverge in the hexagonal structure.

In the LMTO calculations, an extra potential $ \pm \Delta/2$ was added to the on-site energies of the carbon 
orbitals of the two layers of BLG and the band structure was determined self-consistently. The screened potential $\Delta_s$ can then be directly obtained from the final on-site energies of the carbon orbitals by examining the band-center energies. 
For this purpose, the C 2s orbital is especially useful, since it acts like a core orbital. 

For the  TB as well as the analytic calculation, we computed the polarization field $E_P$ by approximating the induced charges in the graphene layer by a uniform sheet charge density $\sigma$, which turns out to be an excellent approximation for computing the screening. The polarization field is then given from the Gauss' Law in electrostatics to be 
$E_p = \sigma / (2\epsilon)$, so that
\begin{equation}
E_s = E - E_p = E - \frac{\delta n}{2\epsilon},
\end{equation}
where $\delta n$ is the difference between the sheet carrier density of the two graphene layers
and $\epsilon$ is the dielectric constant. In terms of the electric
potential the above equation can be wrtitten as
\begin{equation}
\Delta_s = \Delta - \frac{ed\delta n}{2\epsilon},
\label{screeneq}
\end{equation}
where $d$ is the interlayer separation. One can obtain the value of $\delta n$ from the normalized eigenfunctions of the BLG Hamiltonian.  

Diagonalizing the Hamiltonian for the hexagonal BLG (Eq. \ref{HAA}), the eigenfunctions corresponding to the upper and the lower Dirac cones, centered at $\xi$ and $-\xi$
 respectively, are found to be
%
\begin{equation}
\begin{array}{cc}
|\psi_{\pm}^l\rangle =
p_{-}\left(\begin{array}{c}
\pm e^{-i\phi}\\

\mp q_+ e^{-i\phi}\\

-q_+\\

1\\

\end{array}\right),&
|\psi_{\pm}^u\rangle =
p_{+}\left(\begin{array}{c}
\pm e^{-i\phi}\\

\pm q_- e^{-i\phi}\\

q_-\\

1\\
\end{array}\right),
\end{array}
\label{psi}
\end{equation}
where 
$
\phi = \tan^{-1}(k_y/k_x), 
p_{\pm} =  (1\pm \frac{\Delta}{2\xi})^{1/2} / 2,
q_{\pm} = (2t)^{-1} (2\xi \pm \Delta)$,  and
the $\pm$ sign inside the ket  indicates the upper and the lower band and $u (l)$ denotes
the upper (lower)  Dirac cone.
Denoting the four components of the wave function by 
$\psi^k_{A}$, $\psi^k_{\tilde{A}}$, $\psi^k_{\tilde{B}}$, and $\psi^k_{B}$ (from top to bottom), corresponding to the appropriate Bloch functions, the  surface  electron density is given by
\begin{eqnarray}
\sigma^{1(2)} & = & 2e  \sum_{\nu k}^{occ}n_{\nu k}^{1(2)}/A_{cell}, \nonumber \\
n_{\nu k}^{1(2)} & = & |\psi^k_{A(\tilde{A})}|^{2} + |\psi^k_{B(\tilde{B})}|^{2},
\label{n12}
\end{eqnarray}
where A$_{cell}$ is the unit cell area, $\nu$ is the band index, the superscript 1(2) refers to the two individual layers, and the factor of two in front of the summation takes care of the electron spin. 

\begin{figure}
\includegraphics[width=6.0cm]{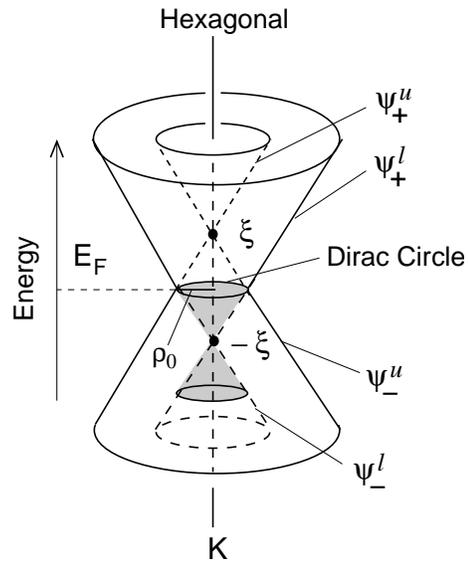}
\caption{The topology of the band structure of hexagonal BLG. The interpenetrating Dirac
cones form a Dirac circle with radius $\rho_0$ that makes the Fermi surface.
The shaded region contributes to the integral in Eq. \ref{screening-integral} in the screening calculation.}
\label{bandsketchaa} 
\end{figure}

The summation in Eq. \ref{n12} need to be performed only over the shaded region in
 Fig. \ref{bandsketchaa}, since we are interested only in the difference $\sigma^1-\sigma^2$.  The remaining portions of the two occupied bands $|\psi_{-}^u\rangle$ and $|\psi_{-}^l\rangle$
   cancel each other's contribution at each $k$ point to this difference as can be easily verified from Eqs. (\ref{psi}) and (\ref{n12}).
    The net result is then
\begin{equation}
\delta n  =  \sigma^{1} - \sigma^{2} = \frac{4e}{\pi}\sum_{\nu = 1}^{2}\int_{0}^{\rho_0}kdk(n_{\nu k}^{1} - n_{\nu k}^{2}),
\label{screening-integral}
\end{equation}
where the summation goes over the two bands below $E_F$ and the integration goes up to the radius of the Dirac circle, so that the contribution comes only from the shaded region of 
 Fig. \ref{bandsketchaa}. 
 The factor of four in Eq. \ref{screening-integral} was inserted to account for both the spin degeneracy and the two Dirac valleys in the Brillouin zone at the K and K$^{\prime}$ points.
Using the wave functions Eq. \ref{psi}, the integration is easily performed to yield 
\begin{equation}
\delta n   =  \frac{e \xi \Delta }{2\pi \nu^{2}}.
\label{delta-n}
\end{equation}

    Now, electrons of course see the screened field, so that the net charge density difference $\delta n$ between the layers is obtained by using the screened potential
    $\Delta_s$ in lieu of $\Delta$ in the expression (\ref{delta-n}). This together with  Eq. \ref{screeneq} yields the ratio between the external potential
 and the screened potential, which is
\begin{equation}
\frac{\Delta}{\Delta_s} = 1 +
\frac{e^{2}d}{4 \pi \epsilon v^2} \sqrt{4t^2+\Delta_{s}^2} . 
\label{Delta-ratio}
\end{equation}

This is plotted in Fig. \ref{Delta-ratio} together with the LMTO result and the TB result. The latter was obtained by a full self-consistent calculation using the nearest-neighbor tight-binding model, from which the charge difference $\delta n$ was computed and used in Eq. \ref{screeneq} to obtain the ratio of   $\Delta / \Delta_s$.
In the LMTO results, all electrons including the carbon $\sigma$ electrons participated in the screening, while in the analytical and TB calculation, which included the effects of only the $\pi$ electrons, we replaced, for simplicity, just the vacuum dielectric constant  for $\epsilon$ appearing in Eq. \ref{Delta-ratio}.

\begin{figure}
\includegraphics[width=8.cm]{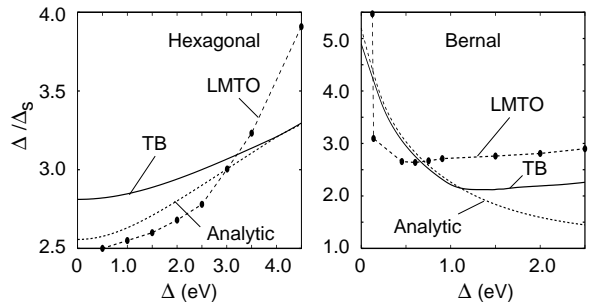}
\caption{\label{screenab} The ratio between the applied electric
potential ($\Delta$) and the screened electric potential ($\Delta_s$) as a function
of $\Delta$ for the hexagonal and the Bernal stacked BLG. Results are shown from the 
analytic calculation, ( Eq. (\ref{Delta-ratio}) for hexagonal and Eq. (\ref{Ratio-Bernal}) for Bernal), the TB calculation, as well as from the density-functional LMTO calculation.
 }
\end{figure}

Fig. \ref{screenab} shows the calculated results for the screening for both the hexagonal and the Bernal structures. For both cases, screening increases with applied electric field, except for the divergence at low field for the Bernal structure as has been pointed out earlier.
The increase of screening with increasing field is due to  the fact that 
proportionately more carriers become involved in screening as the density-of-states for the linear bands near the Fermi energy  increase as square of energy. The analytic and the TB results agree quite well, which is not surprising since the energy spectrum differs only beyond the linear region of the bands and these states don't contribute much because they are far away from the Fermi energy. The density-functional LMTO result, on the other hand, shows larger screening because it includes all carbon electrons that participate in screening.

For the Bernal structure, the divergent screening for low fields may be explained due to the relatively flat bands at the Fermi energy (see Fig. \ref{introfig} a). The effects of the two flat bands touching at the Fermi energy may be taken into account by constructing a $2 \times 2$ effective Hamiltonian using perturbation theory from the full Hamiltonian of Eq. \ref{HAB} and then computing the magnitude of the charge difference from the eigenfuctions. This can be done analytically and, for low fields, the result is\cite{mccann1, min} 
\begin{equation}
\frac{\Delta}{\Delta_{s}} = \frac{2dm}{e^2} \ln \frac{2k_c^2}{ m\Delta_s},
\label{Ratio-Bernal}
\end{equation}
where $m = t_1 (2v^2)^{-1}$ and the cutoff momentum $k_c \approx 0.065$ \AA$^{-1}$. This result has been plotted in Fig. \ref{screenab}
for the Bernal structure and it explains the calculated trend in the density-functional results.

\section{Summary}
In summary, we have studied the effect of the electric field and
strain on the electronic structure of hexagonal and Bernal stacked
bilayer graphene. 
While a gap opens up in the Bernal structure BLG by the application of an external electric field,
the hexagonal BLG remains a zero-gap metal, but with an interesting circular Fermi surface, the Dirac circle, that can be controlled by both strain and electric field. Any doped electrons or holes will occupy a thin shell in the Brillouin zone, which leads to an interesting Fermi surface in the doped system.

This work was supported by the U. S. Department of Energy through Grant No.
DE-FG02-00ER45818.

\end{document}